\documentclass[a4]{article}
\usepackage[T1]{fontenc}
\usepackage[latin1]{inputenc}
\usepackage{amsmath, amssymb, amsthm}
\usepackage[english]{babel}
\usepackage{indentfirst}
\usepackage{booktabs}
\usepackage{array}
\usepackage{caption,appendix}
\usepackage{geometry}
\usepackage{float}
\usepackage{cancel}
\usepackage{bbold}
\usepackage{authblk}
\usepackage{cite}
\usepackage{verbatim}
\usepackage{bm}

\allowdisplaybreaks
\numberwithin{equation}{section}

\theoremstyle{definition}

\begin{document}
\author[1,2,3]{Salvatore Capozziello \thanks{capozziello@na.infn.it}}
\author[4]{Maurizio Capriolo \thanks{mcapriolo@unisa.it} }
\author[5,6]{Shin'ichi Nojiri \thanks{nojiri@gravity.phys.nagoya-u.ac.jp}}
\affil[1]{\emph{Dipartimento di Fisica "E. Pancini", Universit\`a di Napoli {}``Federico II'', Compl. Univ. di
		 Monte S. Angelo, Edificio G, Via Cinthia, I-80126, Napoli, Italy}}
\affil[2]{\emph{INFN Sezione di Napoli, Compl. Univ. di
		 Monte S. Angelo, Edificio G, Via Cinthia, I-80126, Napoli, Italy}}
\affil[3]{\emph{Scuola Superiore Meridionale, Largo S. Marcellino 10, I-80138, Napoli, Italy}}
\affil[4]{\emph{Dipartimento di Fisica Universit\`a di Salerno, via Giovanni Paolo II, 132, Fisciano, SA I-84084, Italy} }
\affil[5]{\emph {Department of Physics,
Nagoya University, Nagoya 464-8602, Japan}}
\affil[6]{\emph {Kobayashi-Maskawa Institute for the Origin of Particles and the Universe, Nagoya University, Nagoya 464-8602, Japan}}
\date{\today}
\title{\textbf{Gravitational waves in $f(Q)$ non-metric gravity via geodesic deviation}}
\maketitle
\begin{abstract} 
We investigate gravitational waves in the $f(Q)$ gravity, i.e., a geometric theory of gravity described by a non-metric compatible connection, free from torsion and curvature, known as symmetric-teleparallel gravity. We show that $f(Q)$ gravity exhibits only two massless and tensor modes. Their polarizations are transverse with helicity equal to two, exactly reproducing the plus and cross tensor modes typical of General Relativity. In order to analyze these gravitational waves, we first obtain the deviation equation of two trajectories followed by nearby freely falling point-like particles and we find it to coincide with the geodesic deviation of General Relativity. This is because the energy-momentum tensor of matter and field equations are Levi-Civita covariantly conserved and, therefore, free structure-less particles follow, also in $f(Q)$ gravity, the General Relativity geodesics. Equivalently, it is possible to show that the curves are solutions of a force equation, where an extra force term of geometric origin, due to non-metricity, modifies the autoparallel curves with respect to the non-metric connection. In summary, gravitational waves produced in non-metricity-based $f(Q)$ gravity behave as those in torsion-based $f(T)$ gravity and it is not possible to distinguish them from those of General Relativity only by wave polarization measurements. This shows that the situation is different with respect to the curvature-based $f(R)$ gravity where an additional scalar mode is always present for $f(R)\neq R$.
\end{abstract}
\section{Introduction}
General Relativity (GR) can be considered the most successful theory of gravitation ever proposed. It has passed several observational probes such as solar system tests and predicted the existence of gravitational waves and black holes, recently observed. However, it is mainly plagued by some critical issues at infrared and ultraviolet scales. In fact, it is difficult to describe the late and early accelerated expansion of the Universe without adding scalar fields or dark components or addressing the problem of the cosmological tensions in the framework of pure GR. From a more theoretical point of view, difficulties in achieving a renormalizable and regularizable behavior of GR at quantum scales prevent the formulation of a self-consistent theory of Quantum Gravity. 
To overcome these problems, modifications, and extensions of GR have been proposed
\cite{Starobinsky, Nojiri1, Caprep, Nojiri2, Clifton, DeFelice, Faraoni, Odi2,Cai}.
The paradigm is to seek for improving or changing the geometric sector in view to achieve a comprehensive theory of gravity working consistently at all scales.

Among these geometric theories, there are those that exploit the possibility of considering more general connections than Levi-Civita (LC) in order to take into account eventual further degrees of freedom for the gravitational field. In this perspective, it is possible to describe spacetime as a differentiable manifold equipped with a general connection where curvature, non-metricity, and torsion can contribute to the dynamics of the gravitational field. That is, metric-affine theories where connection and metric are independent dynamical variables encoding gravity. 

If we constrain the curvature of the connection to vanish, we obtain teleparallel theories~\cite{JHK1} which, in turn, become also symmetric if we turn off the torsion of the connection, thereby obtaining the symmetric teleparallel gravity (STG)~\cite{RV}. 
Remarkable is the so-called geometrical trinity of gravity which provides three equivalent representations of gravitational interaction~\cite{JHK,CDF}: General Relativity, Teleparellel Equivalent of General Relativity (TEGR), and Symmetric Teleparallel Equivalent of General Relativity (STEGR). The first is the standard Einstein theory where spacetime curvature, in the absence of torsion and non-metricity, encodes gravitational field dynamics through the Hilbert-Einstein (GR) action
\begin{equation}\label{01}
S_\mathrm{EH}=\frac{1}{16\pi G}\int_{\Omega}d^{4}x\, \sqrt{-g}\,R \,,
\end{equation} 
where $R$ is the Ricci curvature scalar, $g$ the metric determinant, and $G$ the Newton gravitational coupling.
The second equivalent theory to GR is the teleparallel one, where gravity is described by torsion on a flat spacetime where a compatible metric is assigned. Here the gravitational field is encoded by the torsion scalar $T$ through the TEGR action \cite{Aldrovandi}
\begin{equation}\label{02}
S_\mathrm{TEGR}=\frac{1}{16\pi G}\int_{\Omega}d^{4}x\, e\,T \ ,
\end{equation} 
where $e=\text{det}\,(e^{a}_{\phantom{a}\mu})$, with $\{e^{a}_{\phantom{a}\mu}\}$ is the tetrad basis or {\it vierbeins}, that is an orthogonal local basis of smooth vector fields in Minkowski tangent space to the manifold at a point $\mathcal{P}$, satisfying the relations
\begin{equation}
g_{\mu\nu}=\eta_{ab}\,e^{a}_{\phantom{a}\mu}e^{b}_{\phantom{b}\nu}\ ,
\end{equation} 
with $\eta_{ab}=\text{diag}(1,-1,-1,-1)$ or in terms of the dual basis $\{e_{a}^{\phantom{a}\mu}\}$ satisfying
\begin{equation}
\eta_{ab}=g_{\mu\nu}e_{a}^{\phantom{a}\mu}e_{b}^{\phantom{b}\nu}\ .
\end{equation}
Finally, the third representation, equivalent to GR, is the symmetric teleparallel one. Here gravity is totally described by the non-metricity of flat torsion-free spacetime, and the gravitational field is encoded by the non-metricity scalar $Q$ through the STEGR action
\begin{equation}\label{02_1}
S_\mathrm{STEGR}=\frac{1}{16\pi G}\int_{\Omega}d^{4}x\, \sqrt{-g}\,Q \,,
\end{equation}
provided that we adopt the coincident gauge in which the STG connection can be trivialized
\cite{Sebastian,Jose}.
It is worth noticing that extensions by means of functions of the geometric scalar invariants as $f(R)$, $f(T)$, and $f(Q)$, are not equivalent \cite{CDF}. This fact is a matter of debate because, theories, equivalent to linear actions in the invariants, become different if such linearity is violated. This means that a deep investigation of the number of degrees of freedom of any representation is necessary and this fact could have important consequences in cosmology because it can potentially affect the $\Lambda$CDM model
\cite{Koussour,Mandal,Naik,Vishwakarma,Shi,Ferreira,Bajardi,Koussour1}.

In this paper, we want to search for gravitational waves (GWs) propagating in $f(Q)$ gravity. 
For a recent review on this topic, see Ref.\cite{Lavinia}. Some astrophysical and cosmological applications of $f(Q)$ gravity are reported in \cite{App1,App2,App3,App4,App5,App6,App7}.
With respect to previous studies on GWs in this framework  \cite{HPUS, SFSGS}, we develop our analysis in an alternative way by considering the deviations between nearby trajectories followed by free particles. The equation which we derive reduces to the well-known geodesic deviation in the case of GR.

The first question we have to answer is to realize what kind of curves free particles follow in $f(Q)$ gravity. 
By these curves, it is possible to obtain the deviation between two nearby trajectories. This is the evolution equation of separation vector $\eta^{\alpha}$, which can be used to study polarization and helicity of GWs. 

To answer this question, we need to calculate the LC covariant divergence of matter energy-momentum tensor and then study its conservation. This approach revealed extremely useful in studying other alternative theories of gravity. For example, the energy-momentum tensor of both matter and geometry in non-local theories of gravity has been considered in~\cite{CCL1} while the case of higher-order curvature-based gravity has been studied in~\cite{CCL, CCT, ACCA}. For theories with torsion, see~\cite{Capozziello:2018qcp}. Therefore, after obtaining the field and connection equations by the least action principle, in the coincident gauge, we linearize them and look for wavelike solutions. The properties of GWs will be investigated by the deviation equation. A similar approach in non-local gravity is developed in~\cite{CAPRIOLOM, CCCQG2021, CCN}, while GWs in teleparallel and metric gravity are developed in~\cite{CCC, CCC1, CCC2}. 

The final result is that $f(T)$ and $f(Q)$ gravities are indistinguishable via GWs because both are governed by second-order differential equations, while $f(R)$ theories are described by fourth-order differential equations and exhibit a further scalar mode. This means that $f(T)$, $f(Q)$, and $f(R)$ are different from the GW point of view. On the other hand, $f(Q)$ coincides with GR like $f(T)$ as already demonstrated in \cite{Bamba}. 

The layout of the paper is the following. In Sec.~\ref{A}, we briefly introduce geometrical objects and properties of STG. In Sec.~\ref{B}, we develop the $f(Q)$ gravity, first by deriving the field and connection equations through a variational principle in subsection~\ref{B1}. Subsequently, in Subsections~\ref{B2} and \ref{B2_1}, we study the covariant divergence of the stress-energy tensor with respect to the LC connection and then we obtain the equations of motion and trajectory deviation in $f(Q)$ gravity. Finally, we analyze the GWs in non-metric theory after linearizing the above equations in subsection~\ref{B3}. We look for wave-like solutions and study the wave polarization via geodesic deviation in subsection~\ref{B4} and~\ref{B5}, respectively. Conclusions are drawn in Sec.~\ref{E}.

\section{Geometrical framework in symmetric teleparallel gravity}\label{A}

The most general affine connection $\Gamma^{\alpha}_{\phantom{\alpha}\mu\nu}$ in metric-affine geometry, involving non-vanishing curvature, torsion, and non-metricity, can be uniquely decomposed into three parts as\cite{JHK, DKC, CDF} 
\begin{equation}\label{1}
\Gamma^{\alpha}_{\phantom{\alpha}\mu\nu}=\hat{\Gamma}^{\alpha}_{\phantom{\alpha}\mu\nu}+K^{\alpha}_{\phantom{\alpha}\mu\nu}+L^{\alpha}_{\phantom{\alpha}\mu\nu}\ ,
\end{equation}
where $\hat{\Gamma}^{\alpha}_{\phantom{\alpha}\mu\nu}$ is the LC connection
\begin{equation}\label{2}
\hat{\Gamma}^{\alpha}_{\phantom{\alpha}\mu\nu}:=\frac{1}{2}g^{\alpha\lambda}\left(\partial_{\mu}g_{\lambda\nu}+\partial_{\nu}g_{\lambda\mu}-\partial_{\lambda}g_{\mu\nu}\right)\,.
\end{equation}
$K^{\alpha}_{\phantom{\alpha}\mu\nu}$ is the contorsion tensor defined through the torsion tensor $T^{\alpha}_{\phantom{\alpha}\mu\nu}$
\begin{equation}\label{3}
T^{\alpha}_{\phantom{\alpha}\mu\nu}=2\Gamma^{\alpha}_{\phantom{\alpha}[\mu\nu]}\ ,
\end{equation}
as
\begin{equation}\label{4}
K^{\alpha}_{\phantom{\alpha}\mu\nu}:=\frac{1}{2}g^{\alpha\lambda}\left(T_{\mu\lambda\nu}+T_{\nu\lambda\mu}+T_{\lambda\mu\nu}\right)\ ,
\end{equation}
antisymmetric in the first and third indices 
\begin{equation}\label{5}
K_{\alpha\mu\nu}=-K_{\nu\mu\alpha}\ ,
\end{equation}
whose antisymmetric part is given by
\begin{equation}\label{6}
K^{\alpha}_{\phantom{\alpha}[\mu\nu]}=\frac{1}{2}\Gamma^{\alpha}_{\phantom{\alpha}\mu\nu}\,.
\end{equation}
 $L^{\alpha}_{\phantom{\alpha}\mu\nu}$ is the disformation tensor, symmetric in the second and third indices 
\begin{align}\label{7}
L^{\alpha}_{\phantom{\alpha}\mu\nu}&:=-\frac{1}{2}g^{\alpha\lambda}\left(Q_{\mu\lambda\nu}+Q_{\nu\lambda\mu}-Q_{\lambda\mu\nu}\right)\\
&=\frac{1}{2}Q^{\alpha}_{\phantom{\alpha}\mu\nu}-Q_{(\mu\phantom{\alpha}\nu)}^{\phantom{(\mu}\alpha}\ ,
\end{align}
\begin{equation}\label{8}
L^{\alpha}_{\phantom{\alpha}[\mu\nu]}=0\ ,
\end{equation}
where $Q_{\alpha\mu\nu}$ is the non-metricity tensor defined as 
\begin{equation}\label{9}
Q_{\alpha\mu\nu}=\nabla_{\alpha}g_{\mu\nu}=\partial_{\alpha}g_{\mu\nu}-\Gamma^{\beta}_{\phantom{\beta}\alpha\mu}g_{\beta\nu}-\Gamma^{\beta}_{\phantom{\beta}\alpha\nu}g_{\beta\mu}\ ,
\end{equation}
symmetric in the last two indices 
\begin{equation}\label{10}
Q_{\alpha[\mu\nu]}=0\ .
\end{equation}
STG is a particular non-metric theory of gravity where curvature and torsion of the affine connection vanish, that is,
\begin{equation}\label{11}
R^{\lambda}_{\phantom{\lambda}\mu\nu\sigma}=\Gamma^{\lambda}_{\phantom{\lambda}\mu\sigma,\nu}-\Gamma^{\lambda}_{\phantom{\lambda}\mu\nu,\sigma}+\Gamma^{\beta}_{\phantom{\beta}\mu\sigma}\Gamma^{\lambda}_{\phantom{\lambda}\beta\nu}-\Gamma^{\beta}_{\phantom{\beta}\mu\nu}\Gamma^{\lambda}_{\phantom{\lambda}\beta\sigma}=0\ ,
\end{equation}
and
\begin{equation}\label{12}
T^{\alpha}_{\phantom{\alpha}\mu\nu}=0\ .
\end{equation}
On the other hand, the non-metricity scalar is defined as
\begin{align}\label{14}
Q\equiv&-\frac{1}{4}Q_{\alpha\beta\gamma}Q^{\alpha\beta\gamma}+\frac{1}{2}Q_{\alpha\beta\gamma}Q^{\gamma\beta\alpha}+\frac{1}{4}Q_{\alpha}Q^{\alpha}-\frac{1}{2}Q_{\alpha}\widetilde{Q}^{\alpha}\\
=&\, g^{\mu\nu}\Bigl(L^{\alpha}_{\phantom{\alpha}\beta\mu}L^{\beta}_{\phantom{\beta}\nu\alpha}-L^{\alpha}_{\phantom{\alpha}\beta\alpha}L^{\beta}_{\phantom{\beta}\mu\nu}\Bigr)\ ,
\end{align}
while the two traces of non-metricity tensor are
\begin{equation}\label{15}
Q_{\alpha}\equiv Q_{\alpha\phantom{\mu}\mu}^{\phantom{\alpha}\mu}\ ,
\end{equation}
and
\begin{equation}\label{16}
\widetilde{Q}^{\alpha}\equiv Q_{\mu}^{\phantom{\mu}\mu\alpha}\ .
\end{equation}
Therefore the contractions of the disformation tensor $L^{\lambda}_{\phantom{\lambda}\alpha\beta}$ become
\begin{align}
L^{\lambda}_{\phantom{\lambda}\alpha\lambda}&=-\frac{1}{2}Q_{\alpha}\ ,\label{8.6}\\
L^{\alpha\lambda}_{\phantom{\alpha\lambda}\lambda}&=\frac{1}{2}Q^{\alpha}-\widetilde{Q}^{\alpha}\ .\label{8.7}
\end{align}
Isometry is immediately restored as soon as

\begin{equation}
Q_{\alpha\mu\nu}=\nabla_{\alpha}g_{\mu\nu}=0\,.
\end{equation}
The non-metricity tensor $Q_{\alpha\mu\nu}$ satisfies the following Bianchi identity
\begin{equation}\label{8.8}
\nabla_{[\alpha}Q_{\beta]\mu\nu}=0\ .
\end{equation}
Now, we can introduce a new very useful geometric object, the non-metricity conjugate tensor defined as 
\begin{equation}\label{17}
P^{\alpha}_{\phantom{\alpha}\mu\nu}=\frac{1}{2}\frac{\partial Q}{\partial Q_{\alpha}^{\phantom{\alpha}\mu\nu}}\ ,
\end{equation}
that, according to the property $Q_{\alpha\mu\nu}$~\eqref{10}, assures its symmetry in the last two indices
\begin{equation}
P^{\alpha}_{\phantom{\alpha}\mu\nu}=P^{\alpha}_{\phantom{\alpha}(\mu\nu)}\ .
\end{equation}
This allows us to write the non-metricity scalar Q as
\begin{equation}\label{18}
Q=Q_{\alpha\mu\nu}P^{\alpha\mu\nu}\ .
\end{equation}
The quantity $P^{\alpha}_{\phantom{\alpha}\mu\nu}$ defined in Eq.~\eqref{17}, thanks to the expression of the non-metricity scalar $Q$ given in~\eqref{14}, can be written as 
\begin{eqnarray}\label{24}
P^{\alpha}_{\phantom{\alpha}\mu\nu}&=&\frac{1}{4}\Bigl[-Q^{\alpha}_{\phantom{\alpha}\mu\nu}+2Q_{(\mu\phantom{\alpha}\nu)}^{\phantom{(\mu}\alpha}+Q^{\alpha}g_{\mu\nu}-\widetilde{Q}^{\alpha}g_{\mu\nu}-\delta^{\alpha}_{(\mu}Q_{\nu)}\Bigr]\nonumber\\
&=&-\frac{1}{2}L^{\alpha}_{\phantom{\alpha}\mu\nu}+\frac{1}{4}\bigl(Q^{\alpha}-\widetilde{Q}^{\alpha}\big)g_{\mu\nu}-\frac{1}{4}\delta^{\alpha}_{(\mu}Q_{\nu)}\,.
\end{eqnarray}
It is a sort of superpotential which we will use below.

\section{$f(Q)$ non-metric gravity}\label{B}
Let us consider an extension of the STEGR considering a generic analytic function $f$ of the non-metricity scalar $Q$. The action of $f(Q)$ gravity can be expressed as~\cite{JHKP,DZ} 
\begin{equation}\label{13}
S=\int_{\Omega}d^{4}x\,\Bigl[\frac{1}{2\kappa^{2}}\sqrt{-g}f\left(Q\right)+\lambda_{\alpha}^{\phantom{\alpha}\beta\mu\nu}R^{\alpha}_{\phantom{\alpha}\beta\mu\nu}+\lambda_{\alpha}^{\phantom{\alpha}\mu\nu}T^{\alpha}_{\phantom{\alpha}\mu\nu}+\sqrt{-g}\mathcal{L}_{m}\bigl(g\bigr)\Bigr]\ ,
\end{equation}
where $\kappa^{2}=8\pi G/c^{4}$ and $\lambda_{\alpha}^{\phantom{\alpha}\beta\mu\nu}=\lambda_{\alpha}^{\phantom{\alpha}\beta[\mu\nu]}$, $\lambda_{\alpha}^{\phantom{\alpha}\mu\nu}=\lambda_{\alpha}^{\phantom{\alpha}[\mu\nu]}$ are the Lagrange multipliers needed to impose the two conditions of curvature-free \eqref{11} and torsion-free \eqref{12}.
Here $\mathcal{L}_m$ is the standard matter term.
\subsection{The field and connection equations}\label{B1}
In Palatini approach the metric tensor $g^{\mu\nu}$ and the STG connection $\Gamma^{\mu}_{\phantom{\mu}\alpha\beta}$ are independent variables \cite{Olmo}. So now we vary action~\eqref{13} with respect to the metric tensor $g_{\mu\nu}$ as~\cite{DKC, CDA, BVC, CFSM, VCCE, ABS, Mehdi}
\begin{equation}\label{18.5}
\delta_{g}S=\int_{\Omega}d^{4}x\,\biggl\{\frac{1}{2\kappa^{2}}\Bigl[\delta_{g}\sqrt{-g}f(Q)+\sqrt{-g}\delta_{g}f(Q)\Bigr]+\delta_{g}\Bigl[\sqrt{-g}\mathcal{L}_{m}\Bigr]\biggr\}\ ,
\end{equation}
where
\begin{equation}
\delta_{g}R^{\alpha}_{\phantom{\alpha}\beta\mu\nu}=\delta_{g}T^{\alpha}_{\phantom{\alpha}\mu\nu}=\delta_{g}\lambda_{\alpha}^{\phantom{\alpha}\beta\mu\nu}=\delta_{g}\lambda_{\alpha}^{\phantom{\alpha}\mu\nu}=0\ ,
\end{equation}
because $g^{\mu\nu}$ and $\Gamma^{\mu}_{\phantom{\mu}\alpha\beta}$ are independent. 
The explicit calculation of the variation of $f(Q)$ term in Eq.~\eqref{18.5} yields 
\begin{equation}\label{18.6}
\delta_{g}f(Q)=f_{Q}\Bigl(P_{\mu\alpha\beta}Q_{\nu}^{\phantom{\nu}\alpha\beta}-2Q^{\alpha\beta}_{\phantom{\alpha\beta}\mu}P_{\alpha\beta\nu}\Bigr)\delta g^{\mu\nu}-2f_{Q}P^{\alpha}_{\phantom{\alpha}\mu\nu}\nabla_{\alpha}\delta g^{\mu\nu}\ ,
\end{equation}
where $f_{Q}=\partial f/\partial Q$. Then inserting Eq.~\eqref{18.6} into Eq.~\eqref{18.5}, according to the matter energy-momentum tensor $T_{\mu\nu}$ defined as 
\begin{equation}\label{23}
T_{\mu\nu}=-\frac{2}{\sqrt{-g}}\frac{\delta\bigl(\sqrt{-g}\mathcal{L}_{m}\bigr)}{\delta g^{\mu\nu}}\ ,
\end{equation}
and the variation of the $\sqrt{-g}$ as 
\begin{equation}
\delta \sqrt{-g}=-\frac{1}{2}\sqrt{-g}\,g_{\mu\nu}\delta g^{\mu\nu}\ ,
\end{equation}
Eq.~\eqref{18.5} becomes 
\begin{multline}\label{23.5}
\delta_{g}S=\int_{\Omega}d^{4}x\,\biggl\{\frac{1}{2\kappa^{2}}\biggl[-\frac{1}{2}\sqrt{-g}\,g_{\mu\nu}f\delta g^{\mu\nu}+\sqrt{-g}f_{Q}\Bigl(P_{\mu\alpha\beta}Q_{\nu}^{\phantom{\nu}\alpha\beta}-2Q^{\alpha\beta}_{\phantom{\alpha\beta}\mu}P_{\alpha\beta\nu}\Bigr)\delta g^{\mu\nu}\\
-2\sqrt{-g}f_{Q}P^{\alpha}_{\phantom{\alpha}\mu\nu}\nabla_{\alpha}\delta g^{\mu\nu}\biggr]-\frac{\sqrt{-g}}{2}T_{\mu\nu}\delta g^{\mu\nu}\biggr\}\ .
\end{multline}
Hence, from the following identities 
\begin{equation}
2\sqrt{-g}f_{Q}P^{\alpha}_{\phantom{\alpha}\mu\nu}\nabla_{\alpha}\delta g^{\mu\nu}=\mathcal{D}_{\alpha}\bigl(2f_{Q}\sqrt{-g}P^{\alpha}_{\phantom{\alpha}\mu\nu}\delta g^{\mu\nu}\bigr)-2\nabla_{\alpha}\bigl(\sqrt{-g}f_{Q}P^{\alpha}_{\phantom{\alpha}\mu\nu}\bigr)\delta g^{\mu\nu}\ ,
\end{equation}
where $\mathcal{D}_{\alpha}$ is the LC covariant derivative and 
\begin{equation}
\nabla_{\alpha}\left(\sqrt{-g}A^{\alpha}\right)=\mathcal{D}_{\alpha}\left(\sqrt{-g}A^{\alpha}\right)\ ,
\end{equation}
with a four-vector $A^{\alpha}=2f_{Q}P^{\alpha}_{\phantom{\alpha}\mu\nu}\delta g^{\mu\nu}$, and from the Stokes theorem, the variation~\eqref{23.5} takes the form
\begin{multline}\label{23.6}
\delta_{g}S=\int_{\Omega}d^{4}x\,\biggl\{\frac{1}{2\kappa^{2}}\biggl[-\frac{\sqrt{-g}}{2}\,g_{\mu\nu}f+\sqrt{-g}f_{Q}\Bigl(P_{\mu\alpha\beta}Q_{\nu}^{\phantom{\nu}\alpha\beta}-2Q^{\alpha\beta}_{\phantom{\alpha\beta}\mu}P_{\alpha\beta\nu}\Bigr)\\
+2\nabla_{\alpha}\left(\sqrt{-g}f_{Q}P^{\alpha}_{\phantom{\alpha}\mu\nu}\right)-\frac{\sqrt{-g}}{2}T_{\mu\nu}\biggr]\delta g^{\mu\nu}\biggr\}+\int_{\partial\Omega}dS_{\alpha}\sqrt{-g}A^{\alpha}\ .
\end{multline}
Assuming that the fields on the boundary go to zero, by the principle of least action, $\delta_{g}S=0$, we find the field equations in $f(Q)$ gravity 
\begin{equation}\label{22}
\boxed{
\frac{2}{\sqrt{-g}}\nabla_{\alpha}\left(\sqrt{-g}f_{Q}P^{\alpha}_{\phantom{\alpha}\mu\nu}\right)-\frac{1}{2}g_{\mu\nu}f+f_{Q}\Bigl(P_{\mu\alpha\beta}Q_{\nu}^{\phantom{\nu}\alpha\beta}-2Q^{\alpha\beta}_{\phantom{\alpha\beta}\mu}P_{\alpha\beta\nu}\Bigr)=\kappa^{2}T_{\mu\nu}
}\ ,
\end{equation} 
a system of nonlinear second-order partial differential equations for the metric tensor $g_{\mu\nu}$ and the STG connection $\Gamma^{\alpha}_{\phantom{\alpha}\mu\nu}$.
The $(1,1)$- form of the field Eqs.~\eqref{22} is 
\begin{equation}\label{23.1}
\frac{2}{\sqrt{-g}}\nabla_{\alpha}\left(\sqrt{-g}f_{Q}P^{\alpha\mu}_{\phantom{\alpha\mu}\nu}\right)-\frac{1}{2}\delta^{\mu}_{\phantom{\mu}\nu}f+f_{Q}P^{\mu}_{\phantom{\mu}\alpha\beta}Q_{\nu}^{\phantom{\nu}\alpha\beta}=\kappa^{2}T^{\mu}_{\phantom{\mu}\nu}\ .
\end{equation}
Likewise now, we perform the variation of action~\eqref{13} with respect to the STG affine connection $\Gamma^{\alpha}_{\phantom{\alpha}\mu\nu}$~\cite{JHK1}, i.e.,
\begin{equation}\label{24.1}
\delta_{\Gamma}S=\int_{\Omega}d^{4}x\,\Biggl[\frac{\sqrt{-g}}{2\kappa^{2}}f_{Q}\delta_{\Gamma}Q+\lambda_{\alpha}^{\phantom{\alpha}\beta\mu\nu}\delta_{\Gamma}R^{\alpha}_{\phantom{\alpha}\beta\mu\nu}+\lambda_{\alpha}^{\phantom{\alpha}\mu\nu}\delta_{\Gamma}T^{\alpha}_{\phantom{\alpha}\mu\nu}\Biggr]\ ,
\end{equation}
because 
\begin{equation}\label{24.2}
\delta_{\Gamma}\sqrt{-g}=\delta_{\Gamma}\mathcal{L}_{m}(g)=\delta_{\Gamma}\lambda_{\alpha}^{\phantom{\alpha}\beta\mu\nu}=\delta_{\Gamma}\lambda_{\alpha}^{\phantom{\alpha}\mu\nu}=0\ .
\end{equation}
From the following variations with respect to the STG connection 
\begin{equation}\label{24.3}
\delta_{\Gamma}R^{\alpha}_{\phantom{\alpha}\beta\mu\nu}=2\nabla_{[\mu}\delta\Gamma^{\alpha}_{\phantom{\alpha}\nu]\beta}+T^{\gamma}_{\phantom{\gamma}\mu\nu}\delta\Gamma^{\alpha}_{\phantom{\alpha}\gamma\beta}\ ,
\end{equation}
\begin{equation}\label{24.4}
\delta_{\Gamma}T^{\alpha}_{\phantom{\alpha}\mu\nu}=2\,\delta\Gamma^{\alpha}_{\phantom{\alpha}[\mu\nu]}\ ,
\end{equation}
\begin{equation}\label{24.5}
\delta_{\Gamma}Q=-4\,P^{\mu\nu}_{\phantom{\mu\nu}\alpha}\delta\Gamma^{\alpha}_{\phantom{\alpha}\mu\nu}\ ,
\end{equation}
and taking into account the torsion-free connection and the antisymmetry of multipliers $\lambda_{\alpha}^{\phantom{\alpha}\beta\mu\nu}$ and $\lambda_{\alpha}^{\phantom{\alpha}\beta\mu\nu}$ in their last indices, we have 
\begin{equation}\label{24.6}
\delta_{\Gamma}S=\int_{\Omega}d^{4}x\,\Biggl[-\frac{2\sqrt{-g}}{\kappa^{2}}f_{Q}P^{\mu\nu}_{\phantom{\mu\nu}\alpha}\delta\Gamma^{\alpha}_{\phantom{\alpha}\mu\nu}+2\lambda_{\alpha}^{\phantom{\alpha}\beta\mu\nu}\nabla_{\mu}\delta\Gamma^{\alpha}_{\phantom{\alpha}\nu\beta}+2\lambda_{\alpha}^{\phantom{\alpha}\mu\nu}\delta\Gamma^{\alpha}_{\phantom{\alpha}\mu\nu}\Biggr]\ .
\end{equation}
The second term in the previous integral can be performed in the following form, according to the Lagrange formula and the Stokes theorem
\begin{equation}\label{24.7}
2\int_{\Omega}d^{4}x\,\lambda_{\alpha}^{\phantom{\alpha}\beta\mu\nu}\nabla_{\mu}\delta\Gamma^{\alpha}_{\phantom{\alpha}\nu\beta}=2\int_{\Omega}d^{4}x\,\nabla_{\mu}\lambda_{\alpha}^{\phantom{\alpha}\beta\mu\nu}\delta\Gamma^{\alpha}_{\phantom{\alpha}\nu\beta}-2\int_{\partial\Omega}dS_{\mu}\lambda_{\alpha}^{\phantom{\alpha}\beta\mu\nu}\delta\Gamma^{\alpha}_{\phantom{\alpha}\nu\beta}\ .
\end{equation}
Since we have assumed that the fields on the boundary go to zero, the previous surface integral vanishes. Then, from Eqs.~\eqref{24.6} and \eqref{24.7}, the principle of least action $\delta_{\Gamma}S=0$ leads to the following equation 
\begin{equation}\label{25_1}
\kappa^{2}\Bigl(\nabla_{\beta}\lambda_{\alpha}^{\phantom{\alpha}\nu\mu\beta}+\lambda_{\alpha}^{\phantom{\alpha}\mu\nu}\Bigr)=\sqrt{-g}f_{Q}P^{\mu\nu}_{\phantom{\mu\nu}\alpha}\ .
\end{equation}
Thus, the antisymmetry of the last two indices of the Lagrange multipliers and the commutativity of the STG covariant derivatives, due to the absence of torsion and curvature, yield the connection equation in $f(Q)$ gravity 
\begin{equation}\label{26}
\boxed{
\nabla_{\mu}\nabla_{\nu}\bigl(\sqrt{-g}f_{Q}P^{\mu\nu}_{\phantom{\mu\nu}\alpha}\bigr)=0
}\ .
\end{equation}
It is worth pointing out that in the Palatini approach the metric tensor and the connection are independent variables and then the variation $\delta_{g}$, with respect to $g_{\mu\nu}$ of the connection $\Gamma^{\alpha}_{\phantom{\alpha}\mu\nu}$, vanishes, that is, $\delta_{g}\Gamma^{\alpha}_{\phantom{\alpha}\mu\nu}=0$ and the variation $\delta_{g}$ and covariant derivative $\nabla_{\mu}$ associated to the STG-$\Gamma$ connection commute, that is, 
\begin{equation}
[\nabla_{\mu}, \delta_{g}]A_{\nu}\bigl(g,\Gamma\bigr)=0\ .
\end{equation}
On the other hand, the variation $\delta_{\Gamma}$ and the covariant derivative $\nabla_{\mu}$, associated with the STG-$\Gamma$ connection, do not commute, that is, for a commutator acting on the vector field $A_{\nu}$, one gets
\begin{equation}
[\nabla_{\mu}, \delta_{\Gamma}]A_{\nu}\bigl(g,\Gamma\bigr)=A_{\lambda}\bigl(g,\Gamma\bigr)\delta_{\Gamma}\Gamma^{\lambda}_{\phantom{\lambda}\mu\nu}\ .
\end{equation}
\subsection{The Levi-Civita covariant divergence of the matter energy-momentum tensor}\label{B2}
Now we explicitly calculate the LC covariant divergence of the matter energy-momentum tensor $T^{\mu}_{\phantom{\mu}\nu}$ and verify that it vanishes, that is $\mathcal{D}_{\mu}T^{\mu}_{\phantom{\mu}\nu}=0$. This means that the energy-momentum tensor of the matter is covariantly conserved with respect to the LC connection. To do this, we first multiply both sides of the field Eq.~\eqref{23.1} for $\sqrt{-g}$ and then apply the LC covariant derivative $\mathcal{D}_{\mu}$. Observing that the geometric object $\mathcal{E}^{\mu}_{\phantom{\mu}\nu}=\sqrt{-g}T^{\mu}_{\phantom{\mu}\nu}$, is a mixed tensor density of weight 1 and the quantity $\sqrt{-g}$ is a scalar density of weight $w=1$, than their STG covariant derivative are, respectively~\cite{RV, HKLOR, XLHL} 
\begin{equation}
\nabla_{\alpha}\mathcal{E}^{\alpha}_{\phantom{\alpha}\nu}=\mathcal{D}_{\alpha}\mathcal{E}^{\alpha}_{\phantom{\alpha}\nu}-L^{\lambda}_{\phantom{\lambda}\alpha\nu}\mathcal{E}^{\alpha}_{\phantom{\alpha}\lambda}\ ,
\end{equation}
and 
\begin{equation}\label{8.24}
\nabla_{\alpha}\sqrt{-g}=\,\frac{1}{2}\sqrt{-g}\,Q_{\alpha}\ .
\end{equation}
Then we get
\begin{multline}\label{37.1}
\mathcal{D}_{\mu}\left(\kappa^{2}\sqrt{-g}T^{\mu}_{\phantom{\mu}\nu}\right)=2\nabla_{\mu}\nabla_{\alpha}\left(\sqrt{-g}f_{Q}P^{\alpha\mu}_{\phantom{\alpha\mu}\nu}\right)-\frac{1}{2}\nabla_{\nu}\left(\sqrt{-g}f\right)+\nabla_{\mu}\left(\sqrt{-g}f_{Q}P^{\mu}_{\phantom{\mu}\alpha\beta}Q_{\nu}^{\phantom{nu}\alpha\beta}\right)\\
+2L^{\alpha}_{\phantom{\alpha}\lambda\nu}\nabla_{\mu}\left(\sqrt{-g}f_{Q}P^{\mu\lambda}_{\phantom{\mu\lambda}\alpha}\right)-\frac{1}{2}\sqrt{-g}L^{\alpha}_{\phantom{\alpha}\lambda\nu}\delta^{\lambda}_{\alpha}f+\sqrt{-g}f_{Q}L^{\alpha}_{\phantom{\alpha}\lambda\nu}P^{\lambda}_{\phantom{\lambda}\sigma\eta}Q_{\alpha}^{\phantom{\alpha}\sigma\eta}\ .
\end{multline}
From the connection Eq.~\eqref{26} and the commutativity of the STG covariant derivative, according to Eqs.~\eqref{8.6} and~\eqref{8.24}, and considering that the non-metricity of STG connection yields
\begin{equation}\label{37.2}
\nabla_{\mu}P^{\mu\lambda}_{\phantom{\mu\lambda}\alpha}=g_{\alpha\beta}\nabla_{\mu}P^{\mu\lambda\beta}+Q_{\mu\alpha\beta}P^{\mu\lambda\beta}\ ,
\end{equation}
after some manipulations, Eq.~\eqref{37.1} takes the form
\begin{multline}\label{37.3}
\mathcal{D}_{\mu}\left(\kappa^{2}\sqrt{-g}T^{\mu}_{\phantom{\mu}\nu}\right)=-\frac{1}{2}\sqrt{-g}\nabla_{\nu}f+\nabla_{\mu}\left(\sqrt{-g}f_{Q}\right)\left(Q_{\nu\alpha\beta}+2L_{\beta\alpha\nu}\right)P^{\mu\alpha\beta}\\
+\sqrt{-g}f_{Q}\left(Q_{\nu\alpha\beta}+2L_{\beta\alpha\nu}\right)\nabla_{\mu}P^{\mu\alpha\beta}\\
+\sqrt{-g}f_{Q}\left(\nabla_{\mu}Q_{\nu\alpha\beta}+L^{\rho}_{\phantom{\rho}\mu\nu}Q_{\rho\alpha\beta}+2L^{\rho}_{\phantom{\rho}\alpha\nu}Q_{\mu\beta\rho}\right)P^{\mu\alpha\beta}\ .
\end{multline}
From the symmetry in the last two indices of $P_{\alpha\beta\gamma}$ and the definition of the disformation tensor $L_{\alpha\beta\gamma}$ follow that
\begin{equation}\label{37.4}
\left(Q_{\nu\alpha\beta}+2L_{\beta\alpha\nu}\right)P^{\mu\alpha\beta}=\left(Q_{\nu\alpha\beta}+2L_{\beta\alpha\nu}\right)\nabla_{\mu}P^{\mu\alpha\beta}=0\ .
\end{equation}
Hence, from the Bianchi identities~\eqref{8.8} and the following STG covariant derivative of a generic tensor $F_{\mu\nu\alpha}$ 
\begin{equation}
\nabla_{\beta}F_{\mu\nu\alpha}=\,\mathcal{D}_{\beta}F_{\mu\nu\alpha}-L^{\lambda}_{\phantom{\lambda}\beta\mu}F_{\lambda\nu\alpha}-L^{\lambda}_{\phantom{\lambda}\beta\nu}F_{\mu\lambda\alpha}-L^{\lambda}_{\phantom{\lambda}\beta\alpha}F_{\mu\nu\lambda}\ ,
\end{equation}
joined to Eq.~\eqref{37.4} and the symmetry in the last indices of the $L_{\alpha\beta\gamma}$, Eq.~\eqref{37.3} becomes 
\begin{equation}\label{37.5}
\mathcal{D}_{\mu}\left(\kappa^{2}\sqrt{-g}T^{\mu}_{\phantom{\mu}\nu}\right)=-\frac{1}{2}\sqrt{-g}\nabla_{\nu}f+\sqrt{-g}f_{Q}\left(\mathcal{D}_{\nu}Q_{\mu\alpha\beta}\right)P^{\mu\alpha\beta}\ .
\end{equation}
If we use the definitions of non-metricity conjugate tensor, that is the superpotential $P^{\alpha}_{\phantom{\alpha}\mu\nu}$~\eqref{24} and that of the non-metricity scalar $Q$~\eqref{14}, we have 
\begin{equation}\label{37.6}
Q_{\mu\alpha\beta}\mathcal{D}_{\nu}P^{\mu\alpha\beta}=\frac{1}{2}\mathcal{D}_{\nu}Q\ .
\end{equation}
Finally from the relations~\eqref{37.5} and \eqref{37.6}, we derive that
\begin{equation}\label{37.7}
\mathcal{D}_{\mu}\left(\kappa^{2}\sqrt{-g}T^{\mu}_{\phantom{\mu}\nu}\right)=-\frac{1}{2}\sqrt{-g}\nabla_{\nu}f+\frac{1}{2}\sqrt{-g}f_{Q}\nabla_{\nu}Q=0\ .
\end{equation}
This implies the fundamental result that, in $f(Q)$ gravity, the energy-momentum tensor of matter is covariantly LC conserved or equivalently 
\begin{equation}\label{37.8}
\boxed{
\mathcal{D}_{\mu}T^{\mu}_{\phantom{\mu}\nu}=0
}\,,
\end{equation}
corresponding to the conservation law of energy-momentum tensor of GR, i.e., the contracted Bianchi identities.

\subsection{Equation of motion for a particle and equation of deviation in $f(Q)$ gravity }\label{B2_1}
Eq.~\eqref{37.8} is of crucial importance because it assures that trajectories followed by free point-like particles in $f(Q)$ gravity are the timelike metric geodesics of GR, i.e., 
\begin{equation}\label{51}
\frac{D u^{\lambda}}{ds}=\frac{d^{2}x^{\lambda}}{ds^{2}}+\hat{\Gamma}^{\lambda}_{\phantom{\lambda}\alpha\beta}\frac{dx^{\alpha}}{ds}\frac{dx^{\beta}}{ds}=0\ ,
\end{equation}
where $x^{\mu}(s)$ is a parametric curve with affine parameter $s$, that we identify with the proper time, $u^{\mu}$ the four-velocity tangent to it and $D/ds$ the covariant derivative along our parametric curve expressed as 
\begin{equation}\label{51_11}
\frac{D}{ds}=u^{\alpha}\mathcal{D}_{\alpha}\ .
\end{equation}
Eq.~\eqref{51} can be 
easily obtained if we consider, for the sake of simplicity, dust matter described by the energy-momentum tensor 
\begin{equation}\label{51_25}
T_{\mu\nu}=\rho_{0}u_{\mu}u_{\nu}\ ,
\end{equation}
with $\rho_{0}$ the proper energy density and $u^{\alpha}$ the four-velocity. In this case, we can assume that the pressure is null. From the LC covariant conservation of the stress-energy tensor $T^{\mu}_{\phantom{\mu}\nu}$, we have
\begin{equation}\label{51_30}
\rho_{0}u_{\nu}\mathcal{D}_{\mu}u^{\mu}+\rho_{0}u^{\mu}\mathcal{D}_{\mu}u_{\nu}=0\ .
\end{equation} 
Hence, projecting Eq.~\eqref{51_30} onto the three-space orthogonal to the four-velocity, thanks to the projection operator $h^{\nu}_{\lambda}=\delta^{\nu}_{\lambda}-u_{\lambda}u^{\nu}$ satisfying the orthogonal relation $u_{\nu}h_{\lambda}^{\nu}=0$, we find
\begin{equation}
u_{\lambda}\mathcal{D}_{\mu}u^{\mu}-u_{\lambda}u^{\nu}u_{\nu}\mathcal{D}_{\mu}u^{\mu}+u^{\mu}\mathcal{D}_{\mu}u_{\lambda}-u_{\lambda}u^{\nu}u^{\mu}\mathcal{D}_{\mu}u_{\nu}=0\ .
\end{equation}
Now, we use the normalization condition $u_{\nu}u^{\nu}=1$ and its differentiation $u_{\nu}\mathcal{D}_{\mu}u^{\nu}=0$, so we have the geodesic equation 
\begin{equation}
u^{\mu}\mathcal{D}_{\mu}u^{\lambda}=0\ .
\end{equation}
Geodesic Eq.~\eqref{51} can be equivalently put in the following form where the STG connection $\Gamma^{\lambda}_{\phantom{\lambda}\alpha\beta}$ appears explicitly or via the disformation $L^{\lambda}_{\phantom{\lambda}\alpha\beta}$
\begin{equation}\label{51_20}
\boxed{
\frac{d^{2}x^{\lambda}}{ds^{2}}+\Gamma^{\lambda}_{\phantom{\lambda}\alpha\beta}\frac{dx^{\alpha}}{ds}\frac{dx^{\beta}}{ds}=L^{\lambda}_{\phantom{\lambda}\alpha\beta}\frac{dx^{\alpha}}{ds}\frac{dx^{\beta}}{ds}
}\ ,
\end{equation} 
that is, the equation of the motion of free point-like bodies in $f(Q)$ gravity is a force equation. 
Here it is clear that the free test bodies in $f(Q)$ do not follow autoparallel curves with respect to the STG connection, i.e., those curves along which the tangent vector is propagated parallel to itself with respect to $\Gamma$, because an extra force term appears in the right-hand side of Eq.~\eqref{51_20}, due to the presence of the non-metricity. Eq.~\eqref{51_20} can be put into a more compact and more easily understandable form, provided that we define the covariant derivative along the parametric curve $x^{\mu}(s)$ with respect to STG connection as 
\begin{equation}\label{51_11_1}
\frac{D^{\prime}}{ds}=u^{\alpha}\nabla_{\alpha}\ ,.
\end{equation}
For a arbitrary four-vector $A^{\mu}$, it is related to $D/ds$ as
\begin{equation}
\frac{D^{\prime}A^{\mu}}{ds}=\frac{D A^{\mu}}{ds}+L^{\mu}_{\phantom{\mu}\alpha\beta}u^{\alpha}A^{\beta}\ .
\end{equation}
Finally, Eq.~\eqref{51_20} becomes the following equation of motion in $fQ)$ gravity 
\begin{equation}\label{51_12}
\boxed{
\frac{D^{\prime}u^{\lambda}}{ds}=L^{\lambda}_{\phantom{\lambda}\alpha\beta}\frac{dx^{\alpha}}{ds}\frac{dx^{\beta}}{ds}
}\ .
\end{equation}
So the deviation between two nearby metric geodesics, traveled by two free-falling test bodies in $f(Q)$ gravity, is exactly the geodesic deviation equation of GR, i.e.,
\begin{equation}\label{52_2}
\frac{D^{2}\eta^{\mu}}{ds^{2}}=-\mathcal{R}^{\mu}_{\phantom{\mu}\alpha\nu\beta}\eta^{\nu}u^{\alpha}u^{\beta}\ ,
\end{equation}
where the spacelike separation four-vector $\eta^{\mu}$ connects two nearby points with the same value of $s$ on the two curves. In Eq.~\eqref{52_2}, the Riemann tensor $\mathcal{R}^{\mu}_{\phantom{\mu}\alpha\nu\beta}$ is expressed in terms of the LC connection $\hat{\Gamma}^{\mu}_{\phantom{\lambda}\alpha\beta}$. In other words, Eq.~\eqref{52_2} describes the relative acceleration or the tidal force of neighboring freely falling test bodies. Now we can find the equivalent of the geodesic deviation of GR, in terms of the non-metricity $Q_{\alpha\mu\nu}$ or, more correctly, in terms of disfomation tensor $L^{\alpha}_{\phantom{\alpha}\mu\nu}$. The deviation equation in $f(Q)$ gravity is 
\begin{equation}\label{37.82}
\boxed{
\frac{D^{\prime 2}\eta^{\mu}}{ds^{2}}-2u^{\alpha}L^{\mu}_{\phantom{\mu}\alpha\beta}\frac{D^{\prime}\eta^{\beta}}{ds}=\nabla_{\nu}L^{\mu}_{\phantom{\mu}\alpha\beta}\eta^{\nu}u^{\alpha}u^{\beta}
}\ ,
\end{equation}
where the relative acceleration of two free test bodies is described by the non-metricity of the connection which plays the same role as the GR Riemann tensor. Therefore just as in Riemann spacetime the curvature is probed by the Riemann tensor by the geodesic deviation~\eqref{52_2}, in the symmetric teleparallel spacetime, the non-metricity is probed by the disformation tensor in Eq.~\eqref{37.82}. This is because, in non-metric theories, due to the absence of both torsion and curvature, gravity is encoded by non-metricity. With these considerations in mind, let us seek for GWs in $f(Q)$ gravity.

\subsection{Linearized field equations}\label{B3}
It is well known that in STG, it is always possible to trivialize the connection, that is, we can choose the so-called {\it coincident gauge}~\cite{NF, CONKOI, BJHK} where
\begin{equation}\label{19}
\Gamma^{\alpha}_{\phantom{\alpha}\mu\nu}=0\,,
\end{equation}
imposed to Eq.\eqref{1}.
Indeed, the flatness of connection~\eqref{11} makes it integrable and therefore it can be written as
\begin{equation}
\Gamma^{\alpha}_{\phantom{\alpha}\mu\nu}=\left(\Lambda^{-1}\right)^{\alpha}_{\phantom{\mu}\gamma}\partial_{\mu}\Lambda^{\gamma}_{\phantom{\gamma}\nu}\ ,
\end{equation}
with the matrices $\Lambda^{\gamma}_{\phantom{\gamma}\nu}\in \text{GL}(4,\mathbb{R})$, i.e., belonging to the general linear group. On the other hand, the absence of torsion~\eqref{12} gives 
\begin{equation}
\partial_{[\beta}\Lambda^{\gamma}_{\phantom{\gamma}\nu]}=0\ .
\end{equation}
This implies that the transformation can be parameterized with a vector $\xi^{\mu}$ as
\begin{equation}
\Lambda^{\mu}_{\phantom{\mu}\nu}=\partial_{\nu}\xi^{\mu}\ .
\end{equation}
Finally, if the connection is given by
\begin{equation}
\Gamma^{\alpha}_{\phantom{\alpha}\mu\nu}=\frac{\partial x^{\alpha}}{\partial\xi^{\beta}}\partial_{\mu}\partial_{\nu}\xi^{\beta}\,,
\end{equation}
it is always possible to choose a coordinate system, $\xi^{\mu}=x^{\mu}$, such that the connection vanishes. Hence, in the coincident gauge, we have 
\begin{equation}\label{20}
\hat{\Gamma}^{\alpha}_{\phantom{\alpha}\mu\nu}=-L^{\alpha}_{\phantom{\alpha}\mu\nu}\ ,
\end{equation}
and the non-metricity tensor becomes 
\begin{equation}\label{21}
Q_{\alpha\mu\nu}=\partial_{\alpha}g_{\mu\nu}\ ,
\end{equation}
that is, the covariant derivative $\nabla$ associated to connection $\Gamma$ becomes the partial derivative $\partial$.
It is worth noting that $\nabla_{\alpha}g_{\mu\nu}\neq 0$, while only $\mathcal{D}_{\alpha}g_{\mu\nu}=0$, where $\mathcal{D}_{\alpha}$, as already mentioned, represents the covariant derivative associated to the LC connection $\hat{\Gamma}^{\alpha}_{\phantom{\alpha}\mu\nu}$.
So from now on, we can choose the coincident gauge. 

Let us weakly perturb the metric tensor $g_{\mu\nu}$ around a Minkowsky background as follows
\begin{equation}\label{25}
g_{\mu\nu}=\eta_{\mu\nu}+h_{\mu\nu}\ .
\end{equation}
In coincident gauge, to the first order in the perturbation $h_{\mu\nu}$, the linearized field equations in the vacuum of Eq.~\eqref{22}, setting $T^{\mu}_{\phantom{\mu}\nu}=0$, reduce to~\cite{HPUS, SFSGS} 
\begin{equation}\label{26_5}
f_{Q}(0)\Bigl[\Box h_{\mu\nu}-\bigl(\partial_{\alpha}\partial_{\mu}h^{\alpha}_{\phantom{\alpha}\nu}+\partial_{\alpha}\partial_{\nu}h^{\alpha}_{\phantom{\alpha}\mu}\bigr)-\eta_{\mu\nu}\bigl(\Box h-\partial_{\alpha}\partial_{\beta}h^{\alpha\beta}\bigr)+\partial_{\mu}\partial_{\nu}h\Bigl]=0\ ,
\end{equation}
where $\Box=\eta^{\alpha\beta}\partial_{\alpha}\partial_{\beta}$ is the standard d'Alembert operator. Taking into account the following expansion of $f(Q)$ in terms of non-metricity scalar $Q$
\begin{equation}\label{27}
f(Q)=f(0)+f_{Q}(0)Q+\mathcal{O}(Q^2)\ ,
\end{equation}
the assumption $f(0)=0$ implies 
\begin{align}\label{28}
f(Q)^{(1)}=0\ , \\
f(Q)^{(2)}=f_{Q}(0)Q^{(2)}\ ,
\end{align}
to the first and second order in the expansion parameter $h$. The trace of Eq.~\eqref{26_5}, coinciding with the linearized trace of Eq.~\eqref{22}, becomes 
\begin{equation}\label{29}
\boxed{
\Box h=\partial_{\alpha}\partial_{\beta}h^{\alpha\beta}
}\ ,
\end{equation}
and Eq.~\eqref{26_5} reduces to 
\begin{equation}\label{30}
\boxed{
\Box h-2\partial_{(\mu}\partial_{\alpha}h^{\alpha}_{\phantom{\alpha}\nu)}+\partial_{\mu}\partial_{\nu}h=0
}\ ,
\end{equation}
a system of l second-order linear partial differential equations for the perturbations $h_{\mu\nu}$.

\subsection{Wave solutions}\label{B4}
In order to search for wave solutions of Eqs.~\eqref{30}, it is convenient to use the Fourier transformations.
Then, the field Eqs.~\eqref{30} in the Fourier $\bm{k}$-vector space, according to the following waves expansion 
\begin{equation}\label{64.5}
h_{\mu\nu}(x)=\frac{1}{(2\pi)^{3/2}}\int\,d^{3}k\;\widetilde{h}_{\mu\nu}\bigl(\bm{k}\bigr)e^{ik\cdot x}\ ,
\end{equation}
becomes
\begin{equation}\label{31}
F_{\mu\nu}=k^{2}\tilde{h}_{\mu\nu}-k_{\mu}k^{\alpha}\tilde{h}_{\alpha\nu}-k_{\nu}k^{\alpha}\tilde{h}_{\alpha\mu}+k_{\mu}k_{\nu}\tilde{h}=0\ ,
\end{equation}
while the trace Eq.~\eqref{29} in momentum space reads as
\begin{equation}\label{32}
k^{2}\tilde{h}-k^{\alpha}k^{\beta}\tilde{h}_{\alpha\beta}=0\ .
\end{equation}
Now, we suppose that the wave propagates in $+z$ direction with the wave vector $k^{\mu}=\bigl(\omega,0,0,k_{z}\bigr)$, where $k^{2}=\omega^{2}-k_{z}^{2}$. Thus the wave expansion~\eqref{64.5} reads as 
\begin{equation}\label{43}
h_{\mu\nu}(z,t)=\frac{1}{\sqrt{2\pi}}\int dk_{z}\Bigl(\tilde{h}_{\mu\nu}(k_{z})e^{i(\omega t-k_{z}z)}+c.c.\Bigr)\ ,
\end{equation}
where $c.c.$ stands for complex conjugate. The ten components of linear field Eqs.~\eqref{31}, in the $\bm{k}$-space, assume the following form for a wave propagating along the positive $z$-axis:
\begin{equation}\label{33}
\begin{aligned}
F_{00}&=\bigl(\omega^{2}+k_{z}^{2}\bigr)\tilde{h}_{00}+2\omega k_{z}\tilde{h}_{03}-\omega^{2}\tilde{h}=0\ ,\\
F_{01}&=k_{z}^{2}\tilde{h}_{01}+\omega k_{z}\tilde{h}_{13}=0\ ,\\
F_{02}&=k_{z}^{2}\tilde{h}_{02}+\omega k_{z}\tilde{h}_{23}=0\ ,\\
F_{03}&=\omega k_{z}\bigl(\tilde{h}_{00}-\tilde{h}_{33}-\tilde{h}\bigr)=0\ ,\\
F_{11}&=k^{2}\tilde{h}_{11}=0\ ,\\
F_{12}&=k^{2}\tilde{h}_{12}=0\ ,\\
F_{13}&=\omega k_{z}\tilde{h}_{01}+\omega^{2}\tilde{h}_{13}=0\ ,\\
F_{22}&=k^{2}\tilde{h}_{22}=0\ ,\\
F_{23}&=\omega k_{z}\tilde{h}_{02}+\omega^{2}\tilde{h}_{23}=0\ ,\\
F_{33}&=2\omega k_{z}\tilde{h}_{03}+\bigl(\omega^{2}+k_{z}^{2}\bigr)\tilde{h}_{33}+k_{z}^{2}\tilde{h}=0\,.
\end{aligned}
\end{equation}
The linear trace~\eqref{32}, in the $\bm{k}$-space, becomes
\begin{equation}\label{34}
\omega^{2}\tilde{h}_{00}+2\omega k_{z}\tilde{h}_{03}+k_{z}^{2}\tilde{h}_{33}-\bigl(\omega^{2}-k_{z}^{2}\bigr)\tilde{h}=0\ ,
\end{equation}
where $\tilde{h}$ is the trace of the metric perturbation $h_{\mu\nu}$ in the momentum space given by
\begin{equation}
\tilde{h}=\tilde{h}_{00}-\tilde{h}_{11}-\tilde{h}_{22}-\tilde{h}_{33}\ .
\end{equation}
We first solve the set of Eqs.~\eqref{33} and Eq.~\eqref{34} for $k^{2}\neq 0$. It is straightforward to obtain the following solution
\begin{equation}\label{35}
\begin{aligned}
\tilde{h}_{11}&=\tilde{h}_{12}=\tilde{h}_{22}=0\ ,\\
\tilde{h}_{13}&=-\frac{k_{z}}{\omega}\tilde{h}_{01}\ ,\\
\tilde{h}_{23}&=-\frac{k_{z}}{\omega}\tilde{h}_{02}\ ,\\
\tilde{h}_{33}&=-2\frac{k_{z}}{\omega}\tilde{h}_{03}-\frac{k_{z}^{2}}{\omega^{2}}\tilde{h}_{00}\ ,
\end{aligned}
\end{equation}
with four independent variables $\tilde{h}_{01}$, $\tilde{h}_{02}$, $\tilde{h}_{03}$ and $\tilde{h}_{00}$.
Then, in the case $k^{2}=0$, the solution of Eqs.~\eqref{33} and \eqref{34}, where $\omega=k_{z}$, becomes 
\begin{equation}\label{36}
\begin{aligned}
\tilde{h}_{22}&=-\tilde{h}_{11}\ ,\\
\tilde{h}_{13}&=-\tilde{h}_{01}\ ,\\
\tilde{h}_{23}&=-\tilde{h}_{02}\ ,\\
\tilde{h}_{33}&=-2\tilde{h}_{03}-\tilde{h}_{00}\ ,
\end{aligned}
\end{equation}
with six independent variables $\tilde{h}_{12}$, $\tilde{h}_{11}$, $\tilde{h}_{01}$, $\tilde{h}_{02}$, $\tilde{h}_{03}$ and $\tilde{h}_{00}$.
We will study, in the next subsection, if the found solutions are physical.

\subsection{Polarization via the geodesic deviation}\label{B5}
As a consequence of the vanishing of LC covariant divergence of the matter energy-momentum tensor in $f(Q)$ gravity, i.e., $\mathcal{D}_{\mu}T^{\mu}_{\phantom{\mu}\nu}=0$, we can use the geodesic deviation~\eqref{52_2} or equivalently Eq.~\eqref{37.82} to study the polarization of GWs. Specifically, the solutions of the deviation equation will allow us to understand if solutions~\eqref{35} and \eqref{36} are waves and, if so, what their polarizations are. The displacement $\eta^{\mu}$, which lies on the three-space orthogonal to the four-velocity $u^{\alpha}$, can be chosen as $\eta^{\mu}=(0,\vec{\chi})$ where $\vec{\chi}= (\chi_{x},\chi_{y},\chi_{z})$ is a spatial separation vector that connects two neighboring particles with non-relativistic velocity at rest in the free-falling local frame. We set up a quasi-Lorentz, normal coordinate system with origin on one particle and spatial coordinate $\chi^{i}$ for the other. The spatial components of the geodesic deviation~\eqref{52_2} to the first order in metric perturbation $h_{\mu\nu}$ are~\cite{CMW} 
\begin{equation}\label{40}
\ddot{\chi}^{i}=-\mathcal{R}^{i(1)}_{\phantom{i}0j0}\chi^{j}\ ,
\end{equation}
where $i,j$ range over $(1,2,3)$, and the electric components of the linearized Riemann tensor $\mathcal{R}^{\alpha}_{\phantom{\alpha}\beta\mu\nu}$ are
\begin{equation}\label{41}
\mathcal{R}^{(1)}_{i0j0}=\frac{1}{2}\bigl(h_{i0,0j}+h_{0j,i0}-h_{ij,00}-h_{00,ij}\bigr)\ .
\end{equation}
The Eq.~\eqref{40} can be regarded as the relative acceleration between the two freely falling point-like particles. Inserting Eq.~\eqref{41} into Eq.~\eqref{40}, the linear system of differential equations, for a wave traveling along positive $z$-axis in a proper local reference frame, reads as
\begin{equation}\label{42}
\left\{
\begin{array}{lr}
\ddot{\chi}_{x}=-\frac{1}{2}h_{11,00}\chi_{x}-\frac{1}{2}h_{12,00}\chi_{y}+\frac{1}{2}\bigl(h_{01,03}-h_{13,00}\bigr)\chi_{z}\\
\ddot{\chi}_{y}=-\frac{1}{2}h_{12,00}\chi_{x}-\frac{1}{2}h_{22,00}\chi_{y}+\frac{1}{2}\bigl(h_{02,03}-h_{23,00}\bigr)\chi_{z}\\
\ddot{\chi}_{z}=\frac{1}{2}\bigl(h_{01,03}-h_{13,00}\bigr)\chi_{x}+\frac{1}{2}\bigl(h_{02,03}-h_{23,00}\bigr)\chi_{y}+\frac{1}{2}\bigl(2h_{03,03}-h_{33,00}-h_{00,33}\bigr)\chi_{z}
\end{array}\,.
\right.
\end{equation}
In the case $k^{2}=M^{2}\neq 0$, we insert the solution~\eqref{35} in the system~\eqref{42}, keeping $k_{z}$ fixed. It takes the form
\begin{equation}\label{44}
\left\{
\begin{array}{lr}
\ddot{\chi}_{x}=0\\
\ddot{\chi}_{y}=0\\
\ddot{\chi}_{z}=0
\end{array}\ .
\right.
\end{equation}
Also, imposing the initial conditions, the initial displacement $\bm{\chi}(0)=\bm{R}=(\chi_{x}^{0},\chi_{y}^{0},\chi_{z}^{0})$ and the initial relative velocity $\dot{\bm{\chi}}(0)=0$, after double integration with respect to $t$ of the the system~\eqref{44} we obtain the solution
\begin{equation}\label{45}
\chi_{x}(t)=\chi_{x}^{0}\ ,\quad \chi_{y}(t)=\chi_{y}^{0}\ ,\quad \chi_{z}(t)=\chi_{z}^{0}\ ,
\end{equation}
which is not a wave, that is, there is no mode associated with $k^{2}\neq 0$. That is, when $k^{2}\neq 0$, none of the four degrees of freedom has a physical origin. Instead, in the case $k^{2}=0$, that implies $\omega=k_{z}$, by means of solution~\eqref{36}, the geodesic deviation, at first order in perturbation~\eqref{42} for a fixed $k_{z}$ of a single plane wave, yields 
\begin{equation}\label{46}
\left\{
\begin{array}{lr}
\ddot{\chi}_{x}=\frac{1}{2}\omega^{2}\bigl(\tilde{h}^{(+)}\chi_{x}^{0}+\tilde{h}^{(\times)}\chi_{y}^{0}\bigr)e^{i\omega(t-z)}\\
\ddot{\chi}_{y}=\frac{1}{2}\omega^{2}\bigl(\tilde{h}^{(\times)}\chi_{x}^{0}-\tilde{h}^{(+)}\chi_{y}^{0}\bigr)e^{i\omega(t-z)}\\
\ddot{\chi}_{z}=0
\end{array}\ ,
\right.
\end{equation}
where $\tilde{h}_{11}=\tilde{h}^{(+)}$ and $\tilde{h}_{12}=\tilde{h}^{(\times)}$. Hence, in the coincident gauge, when $k^2=0$, only two degrees of freedom of the initial six survive exactly as in the case of GR. Then, after double integration with respect to $t$, the solution of system~\eqref{46} becomes
\begin{equation}\label{47}
\left\{
\begin{array}{lr}
\chi_{x}(t)=\chi_{x}^{0}-\frac{1}{2}\bigl(\tilde{h}^{(+)}\chi_{x}^{0}+\tilde{h}^{(\times)}\chi_{y}^{0}\bigr)e^{i\omega(t-z)}\\
\chi_{y}(t)=\chi_{y}^{0}-\frac{1}{2}\bigl(\tilde{h}^{(\times)}\chi_{x}^{0}-\tilde{h}^{(+)}\chi_{y}^{0}\bigr)e^{i\omega(t-z)}\\
\chi_{z}(t)=\chi_{z}^{0}
\end{array}\ ,
\right.
\end{equation}
that describes the response of a ring of masses hit by a gravitational wave. When we have the case $\tilde{h}^{(\times)}=0$, from the solution~\eqref{47}, the effect of gravitational wave is to distort the circle of particles into ellipses oscillating in a $+$ pattern, while in the case $\tilde{h}^{(+)}=0$, the ring distorts into ellipses oscillating in a $\times$ pattern rotated by $45$ degrees in a right-handed sense with respect to it. In summary, we have obtained the well-known plus and cross, massless, transverse, and spin 2 modes of GR. The gravitational wave can be put into the form
\begin{equation}\label{48}
h_{\mu\nu}(z,t)=\frac{1}{\sqrt{2\pi}}\int dk_{z}\Bigl[\epsilon_{\mu\nu}^{(+)}\tilde{h}^{(+)}(\omega)+\epsilon_{\mu\nu}^{(\times)}\tilde{h}^{(\times)}(\omega)\Bigr]e^{i\omega (t-z)}+c.c.\ ,
\end{equation}
where $\epsilon_{\mu\nu}^{(+)}$ and $\epsilon_{\mu\nu}^{(\times)}$ are the polarization tensors defined as 
\begin{equation}\label{49}
\epsilon^{(+)}_{\mu\nu}=\frac{1}{\sqrt{2}}
\begin{pmatrix} 
0 & 0 & 0 & 0 \\
0 & 1 & 0 & 0 \\
0 & 0 & -1 & 0 \\
0 & 0 & 0 & 0
\end{pmatrix}\ .
\end{equation}
\begin{equation}\label{49.1}
\epsilon^{(\times)}_{\mu\nu}=\frac{1}{\sqrt{2}}
\begin{pmatrix} 
0 & 0 & 0 & 0 \\
0 & 0 & 1 & 0 \\
0 & 1 & 0 & 0 \\
0 & 0 & 0 & 0
\end{pmatrix}\ .
\end{equation}
Hence if we adopt the coincident gauge in $f(Q)$ non-metric gravity, we obtain exactly the same GWs predicted by GR in the TT gauge. Finally a gravitational wave propagating along an arbitrary $\bm{k}$ direction in $f(Q)$ gravity, from Eq.~\eqref{48}, becomes
\begin{equation}\label{49.2}
\boxed{
h_{\mu\nu}(x)=\frac{1}{(2\pi)^{3/2}}\int d^3k\Bigl[\epsilon_{\mu\nu}^{(+)}\tilde{h}^{(+)}(\bm{k})+\epsilon_{\mu\nu}^{(\times)}\tilde{h}^{(\times)}(\bm{k})\Bigr]e^{i(\omega t-\bm{k}\cdot \bm{x})}+c.c.
}\ .
\end{equation}
\section{Discussion and Conclusions}\label{E}
In this paper, we have presented an approach to analyze the existence and the polarization of GWs in non-metric, torsion-free, and curvature-free theories of gravity, i.e., STG, described by analytic functions of non-metricity scalar $Q$, i.e., $f(Q)$, by means of the geodesic deviation equation. To this aim, we first studied the motion of free point-like particles in our non-metric theory and deduced that they follow ordinary timelike geodesics of GR because the energy-momentum tensor of matter is covariantly conserved with respect to the LC connection, i.e., $\mathcal{D}_{\alpha}T^{\alpha}_{\phantom{\alpha}\beta}=0$. Without any assumption in the Palatini approach, the field and connection equations, obtained by imposing the cancellation of the variations with respect to the metric and the connection, lead to the conservation of the $T^{\alpha}_{\phantom{\alpha}\beta}$ and therefore to the geodesic equation of GR
\begin{equation}\label{53}
\frac{D u^{\lambda}}{ds}=0\ .
\end{equation}
In $f(Q)$ gravity, the geodesic deviation equation can be rewritten in terms of non-metricity as 
\begin{equation}\label{54}
\frac{D^{\prime} u^{\lambda}}{ds}=L^{\lambda}_{\phantom{\lambda}\alpha\beta}\frac{dx^{\alpha}}{ds}\frac{dx^{\beta}}{ds}\ .
\end{equation}
That is, the worldlines of the free structure-less bodies in $f(Q)$ gravity do not follow the autoparallel curves with respect to STC connection, but follow the force Eq.~\eqref{51_12} or equivalently follow the autoparallel curves with respect to LC connection of GR~\eqref{53}. Then the deviation equations for nearby spacetime trajectories are obtained and they are the same as in GR. In other words, it is possible to derive the deviation equation between two nearby curves, traveled by free particles in $f(Q)$: It is Eq.\eqref{37.82} which is equivalent to the geodesic deviation of GR, where the dependence on non-metricity and disformation tensor are explicit. 

In this framework, we can adopt the coincident gauge. It is always possible because torsion and curvature are absent. After we linearized the field equations, we searched for wave solutions via the Fourier transform. Thanks to the geodesic deviation, we verified that the condition $k^{2}\neq 0$ does not give any physical solution while $k^{2}=0$ gives physically viable solutions. Specifically, the solutions of the deviation equation tell us that GWs propagating on the spacetime described by $f(Q)$ gravity are massless tensors with transverse polarization and helicity 2. Furthermore, the two transverse polarizations are exactly the plus and cross tensor modes predicted by GR. So the non-metricity-based $f(Q)$ gravity behaves exactly as torsion-based $f(T)$ gravity and it is not possible to distinguish them from GR only through wave polarization measurements as demonstrated also in \cite{Bamba}.

On the other hand, the curvature-based $f(R)$ exhibits an additional measurable scalar mode. It is possible to show that, if a boundary term is added in $f(Q)$ gravity, the theory can be recast as $f(R)$. See \cite{Carmen} for details.

Specifically,  the further scalar mode of $f(R)$ gravity appears from
the scaling the metric, that is, by a scale transformation, $f(R)$ gravity can be
rewritten as a scalar-tensor theory,  as it is well known. See e.g. \cite{Nojiri1,Caprep,Nojiri2, Clifton, DeFelice}. In \cite{Taishi},   the same procedure has been considered for  $f(Q)$ gravity.  In this case, the corresponding scalar mode appears as a ghost, that is, with a wrong signature of the kinetic term. The scalar mode can be, however, consistently
eliminated by the constraint, and therefore the scalar mode does not propagate.

However, the debate on the number of degrees of freedom, and then the effective modes, in  STG theories is still an open issue. 
In Ref. \cite{Hu:2022anq}, authors claimed that the degrees of freedom in $f(Q)$ gravity are 8, while, in \cite{DAmbrosio:2023asf}, it is reported that the degrees of
freedom should be equal to or less than 7. This means that the question is open and, very likely, it should be addressed by changing some point of view. In particular, in the analog case of $f(T)$ gravity, it is well known that there appear
unphysical (superluminal) modes in higher order perturbations, as reported in 
\cite{Ong:2013qja} and \cite{Izumi:2012qj}. 
Even in the case of $f(Q)$ gravity, there might occur such a thing, so the investigation has to be improved considering also higher-order perturbations.

An important remark is in order at this point. In view to discriminate among concurring theories of gravity at fundamental level, one of the main option is to detect further polarizations with respect to GR. 
 Currently, the only available experimental data are those  from LIGO-Virgo collaboration, which,  unfortunately,  are unable to  distinguish between polarization modes, and cannot precisely detect the energy amount  of  tensor polarization modes. This means that the possibility to discriminate between GR modes and possible further polarizations requires improvements in sensitivities like those of forthcoming   LISA \cite{LISA} and Einstein Telescope (ET) \cite{ET}. For the case presented here, i.e. GWs related to the the geodesic deviation, hopefully, 
combined observations from LISA and ET could distinguish
polarization modes by  detecting scalar modes and even distinguish between the tensor
and, if existing, circular polarizations. An interesting perspective  could be that STG theories could  lead to an overall amplification of the stochastic spectrum, as discussed for   other modified theories of gravity  reported in Ref.\cite{Odi1}. 
This argument and eventual comparison with available data will be the topic of a forthcoming paper.

 \section*{Acknowledgements}
SC acknowledges the Kobayashi-Maskawa Institute (Nagoya), where this study has been developed, for its kind hospitality.
SC and MC acknowledge the Istituto Nazionale di Fisica Nucleare (INFN) Sez. di Napoli, Iniziative Specifiche QGSKY and MOONLIGHT2, and the Istituto Nazionale di Alta Matematica (INdAM), gruppo GNFM, for the support.
This paper is based upon work from COST Action CA21136 {\it Addressing observational tensions in cosmology with systematics and fundamental physics} (CosmoVerse) supported by COST (European Cooperation in Science and Technology).

\end{document}